\documentclass{article}
\usepackage{amsfonts}

\usepackage{graphicx}
\usepackage{amsmath}

\input{tcilatex}

\begin{document}

\title{Feynman Diagrams and the Quantum Stochastic Calculus}
\author{John Gough \\
Department of Computing \& Mathematics\\
Nottingham-Trent University, Burton Street,\\
Nottingham NG1\ 4BU, United Kingdom.\\
john.gough@ntu.ac.uk}
\date{}
\maketitle

\begin{abstract}
We present quantum stochastic calculus in terms of diagrams taking weights
in the algebra of observables of some quantum system. In particular, we note
the absence of non-time-consecutive Goldstone diagrams. We review recent
results in Markovian limits in these terms.
\end{abstract}


\section{Introduction}

Quantum stochastic calculus (QSC) involves an analysis of the fundamental
quantum processes of creation/annihilation/conservation \cite{HP} and
intuitively this is somehow related to emission/absorption/scattering of
physical quanta as described by quantum field theory (QFT). Quantum
stochastic theory has the advantages, as well as the limitations, of having
a mathematically rigorous setting. It also has the theory of classical
probability to fall back on for much of its inspiration. So much so that the
relationship with QFT, which was originally a major motivating factor, is
now frequently overlooked. Effectively, the fundamental quantum processes
should be idealizations of quantum fields for some suitable ``Markovian''
regime. They were introduced to describe open systems dynamics: here the
quantum noise couples to some quantum system and so, in some sense, we are
dealing not just with traditional quantum fields, but with quantum fields
taking values in the algebra of observables of some quantum system.

When presenting his famous list of problems, Hilbert is supposed to have
quoted an unnamed colleague as saying that ``a mathematical theory should
not be considered complete until one can walk out into the street and
explain it to the first person you meet''. Let us suppose we did this and,
as luck would have it, the first person we meet is a physicist. Would we
succeed in explaining quantum stochastic calculus? My contention is that we
should, though we might have to make do with some formal mathematics
(presumably Hilbert wouldn't have objected?). There are many fundamental
ideas, familiar to physicists, hidden (sometimes too well-hidden!) in the
mathematical formalism of quantum stochastic calculus. In this article, I've
tried to present some basic results from quantum stochastic calculus in the
language of QFT and, in particular, in terms Feynman-type diagrams: the hope
being that mathematicians and physicists will learn something from the
cross-over. Diagrammatic techniques are central to QFT \cite{Mattuck}\cite
{AGD}, yet also of independent interest from it; they still remain an
essential tool for investigating mathematical aspects of quantum theory and
continue to yield some of the most illuminating insights (see, for instance,
the papers \cite{BaezDolan}\cite{Gottschalk}\cite{Yoshida}). I also want to
present an account of a recent paper \cite{Gough7} which deals with the QSC
approximation and which was originally formulated in diagrammatic language.

\subsection{Expansions of Evolutions}

A free-particle (Bosonic) quantum field $\Phi _{t}$ living on a Hilbert
(Fock) space $\frak{H}$ can be decomposed into positive and negative
frequency terms as $\Phi _{t}=\Phi _{t}^{\left( +\right) }+\Phi _{t}^{\left(
-\right) }$: here we suppress all dependence other than time and understand
the time label to refer to Heisenberg picture of the free dynamics. We take $%
\Phi _{t}^{\left( +\right) }$ to be an annihilation field and $\Phi
_{t}^{\left( -\right) }$ to be a creation field. If $\Omega $ is the Fock
vacuum vector, then we have the identity 
\begin{equation}
\Phi _{t}^{\left( +\right) }\,\Omega =0,
\end{equation}
along with the canonical commutation relations 
\begin{equation}
\left[ \Phi _{t}^{\left( +\right) },\Phi _{s}^{\left( -\right) }\right]
=G\left( t,s\right) .
\end{equation}
Here $G\left( t,s\right) \equiv \left\langle \Omega |\Phi _{t}\Phi
_{s}\,\Omega \right\rangle $. Related to this is the propagator $K$\ defined
as 
\begin{equation*}
K\left( t,s\right) =\left\langle \Omega |\vec{T}\Phi _{t}\Phi _{s}\,\Omega
\right\rangle =G\left( t,s\right) \theta \left( t-s\right) +G\left(
s,t\right) \theta \left( s-t\right) .
\end{equation*}
As usual $\vec{T}$ is Dyson's chronological operation placing Heisenberg
picture operators \ in increasing time-order from right to left.

Let $\left\{ \Upsilon _{t}:t\geq 0\right\} $ be a family of self-adjoint
operators on $\frak{H}$ with $\Upsilon _{t}$ being some function of $\Phi
_{t}^{\left( \pm \right) }$ We are then interested in the evolution operator 
\begin{equation}
U_{t}=\vec{T}\left\{ \exp -i\int_{0}^{t}\Upsilon _{s}ds\right\} ,
\label{time-ordered U}
\end{equation}
by which we mean the solution to the Schr\"{o}dinger equation $i\partial
_{t}U_{t}=\Upsilon _{t}U_{t},$ $U_{0}=1$. Suppose that we have a polynomial
interaction 
\begin{equation*}
\Upsilon _{t}=\sum_{\nu }\frac{1}{\nu !}\lambda _{\nu }\left( \Phi
_{t}\right) ^{\nu }.
\end{equation*}
The standard device of quantum field theory is to expand $U_{t}$ in terms of
diagrams, see, e.g. \cite{Mattuck} or \cite{Ticciati}. A \textit{Wick diagram%
} $D$ is constructed as follows: choose $n=n\left( D\right) $ labelled
points (vertices), each vertex will have some labelled legs attached (we let 
$m_{\nu }=m_{\nu }\left( D\right) $ denote the number of vertices
having\thinspace $\nu $ legs so that $n=\sum_{\nu }m_{\nu }$), we join
several pairs of legs to form (undirected) edges, the result is a graph
having several external lines and we now ignore the labelling. A Wick
diagram is then the class of all topological equivalent graphs. We let $%
c=c\left( D\right) $ denote the number of ways we could have originally
connected the various legs to get the same graph. Now denote by $\mathcal{D}%
_{W}$ the set of all Wick diagrams and define, for each $D\in \mathcal{D}%
_{W} $, the operator 
\begin{equation}
\tilde{D}\left( t\right) =\left( -i\right) ^{n}\frac{c}{n!}\prod_{\nu
}\left( \frac{\lambda _{\nu }}{\nu !}\right) ^{m_{\nu }}\vec{N}\int_{\left[
0,t\right] ^{n}}\prod_{D}K\prod_{D}\left( \Phi ^{\left( +\right) }+\Phi
^{\left( -\right) }\right)  \label{Wick-term}
\end{equation}
where $\vec{N}$ is normal ordering (placing all creation fields $\Phi
^{\left( -\right) }$ to the left of all annihilation fields $\Phi ^{\left(
-\right) }$) and under the integral we have a factor $K\left(
t_{i},t_{j}\right) $ for each edge $\left( i,j\right) $ occurring and a
factor $\Phi _{t_{k}}$ for each external line at vertex $\left( k\right) $.
It is then a basic result of QFT that $U_{t}$ admits the expansion 
\begin{equation*}
U_{t}=\sum_{D\in \mathcal{D}_{W}}\tilde{D}\left( t\right) .
\end{equation*}
Next let $\mathcal{P}_{W}$ denote the subset of \textit{connected} Wick
diagrams then we may list the elements as $P_{1},P_{2},\cdots $ and each $%
D\in \mathcal{D}_{W}$ can be decomposed as $D\equiv P_{1}^{n_{1}}\times
P_{2}^{n_{2}}\times \cdots $. Now one readily checks that $\tilde{D}\left(
t\right) =\vec{N}\dfrac{\tilde{P}_{1}\left( t\right) ^{n_{1}}}{n_{1}!}\dfrac{%
\tilde{P}_{2}\left( t\right) ^{n_{2}}}{n_{2}!}\cdots $ and so 
\begin{equation}
U_{t}=\vec{N}\sum_{n_{1},n_{2},\cdots }\prod_{j=0}^{\infty }\dfrac{\tilde{P}%
_{j}\left( t\right) ^{n_{j}}}{n_{j}!}=\vec{N}\prod_{j=0}^{\infty
}\sum_{n=0}^{\infty }\dfrac{\tilde{P}_{j}\left( t\right) ^{n_{j}}}{n_{j}!}=%
\vec{N}\exp \sum_{P\in \mathcal{P}_{W}}\tilde{P}\left( t\right) .
\label{sum-over-connected-Wick-diagrams}
\end{equation}
What we have managed to do is to express the evolution operator $U_{t}$ as a
normal ordered exponential of a sum over connected Wick diagrams. The
connected Wick diagrams play the role of the `primes' amongst the set of all
Wick diagrams - indeed the trick of replacing a sum of products by a product
of sums is just the one that goes on when we develop a prime number
expansion of the Dirichlet series of a multiplicative function, the Riemann
zeta function being perhaps the best known example; it is also the trick
used to compute the grand canonical partition function for the free Bose
gas. The result should be understood as an operator theoretic version of the
usual cumulant moment expansion.

\bigskip

Now let $\frak{h}_{S}$\ be a fixed Hilbert space. We move the action up to
the Hilbert space $\frak{h}\otimes \frak{H}$ and set 
\begin{equation}
\Upsilon _{t}=\sum_{\alpha ,\beta }E_{\alpha \beta }\otimes \left( \Phi
_{t}^{\left( -\right) }\right) ^{\alpha }\left( \Phi _{t}^{\left( +\right)
}\right) ^{\beta }  \label{interaction}
\end{equation}
where we take $E_{\alpha \beta }^{\dagger }=E_{\beta \alpha }$. We now
introduce a class of diagrams known as \textit{Goldstone diagrams} - they
differ from the previous ones in that the vertices are placed in time order 
\cite{Mattuck}. Consider times $t_{n}>\cdots >t_{2}>t_{1}$ in the interval $%
\left[ 0,t\right] $ and draw these as vertices as shown below:

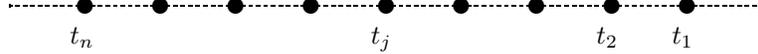
\begin{center}
%
%
%
%
%
%
\setlength{\unitlength}{.1cm}
\begin{picture}(120,20)
\label{pic1}

\put(20,10){\circle*{2}}
\put(30,10){\circle*{2}}
\put(40,10){\circle*{2}}
\put(50,10){\circle*{2}}
\put(60,10){\circle*{2}}
\put(70,10){\circle*{2}}
\put(80,10){\circle*{2}}
\put(90,10){\circle*{2}}
\put(100,10){\circle*{2}}

\put(18,5){$t_{n}$}
\put(88,5){$t_{2}$}
\put(98,5){$t_{1}$}
\put(58,5){$t_{j}$}

\put(10,10){\dashbox{0.5}(100,0){ }}

\end{picture}
%
.
\end{center}

Suppose that at vertex $j$ we have $\beta _{j}$ legs coming in from the
right, representing annihilators, and $\alpha _{j}$ legs going out to the
left, representing creators. For example, the vertex for $E_{23}\otimes %
\left[ \Phi _{t_{j}}^{\left( -\right) }\right] ^{2}\left[ \Phi
_{t_{j}}^{\left( +\right) }\right] ^{3}$ where we have $\alpha _{j}=2$
creators and $\beta _{j}=3$ annihilators is sketched as

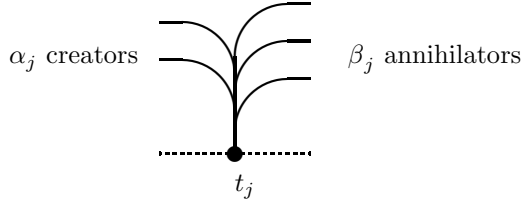
\begin{center}
%
%
%
%
%
\setlength{\unitlength}{.1cm}
\begin{picture}(40,30)
\label{pic2}

\put(10,8){\dashbox{0.5}(20,0){ }}

\thicklines

\put(20,8){\circle*{2}}
\put(30,8){\oval(20,20)[tl]}
\put(30,8){\oval(20,40)[tl]}
\put(30,8){\oval(20,30)[tl]}
\put(10,8){\oval(20,25)[tr]}
\put(10,8){\oval(20,35)[tr]}
\put(20,3){$t_{j}$}
\put(-10,20){$\alpha_{j}$ creators}
\put(35,20){$\beta_{j}$ annihilators}

\end{picture}
%
.
\end{center}

We construct a Goldstone diagram $D$ as follows. We take an arbitrary number 
$n=n\left( D\right) $ vertices and draw in an ordered line as above. We then
draw creation / annihilation lines at each vertex corresponding to one of
the terms appearing in $\Upsilon $. We then connect selected creation legs
to (necessarily later time) annihilation legs: the remaining uncontracted
legs are then \textit{directed} external lines. We consider the family $%
\mathcal{D}_{G}$ of all (topologically distinct) diagrams obtained in this
way. To each Goldstone diagram $D$ we associate the operator 
\begin{equation}
\hat{D}\left( t\right) =\left( -i\right) ^{n}E_{\alpha _{n}\beta _{n}}\cdots
E_{\alpha _{1}\beta _{1}}\otimes \int_{\Delta _{n}\left( t\right)
}\prod_{D}\Phi ^{\left( -\right) }\prod_{D}G\prod_{D}\Phi ^{\left( +\right) }
\label{Goldstone-term}
\end{equation}
where $\Delta _{n}\left( t\right) $ is the simplicial region $\left\{ \left(
t_{n},\cdots ,t_{1}\right) :t>t_{n}>\cdots >t_{2}>t_{1}>0\right\} $ and we
have a factor $G\left( t_{i}-t_{j}\right) $ for each edge $\left( i,j\right) 
$, note $t_{i}>t_{j}$, a factor $\Phi _{t_{k}}^{\left( +\right) }$ for each
incoming external line to a vertex $k$ and a factor $\Phi _{t_{k}}^{\left(
-\right) }$ for each outgoing external line. We then find the expansion 
\begin{equation}
U_{t}=\sum_{D\in \mathcal{D}_{G}}\hat{D}\left( t\right) .
\label{sum-over-Goldstone-diagrams}
\end{equation}

In QFT one is used to switching between an expansion in terms of Goldstone
diagrams $\left( \ref{sum-over-Goldstone-diagrams}\right) $\ and one in
terms of Wick diagrams $\left( \ref{sum-over-connected-Wick-diagrams}\right) 
$. In the present case however we have an obstruction: the $E_{\alpha \beta
} $'s do not necessarily commute! This complication means that the Goldstone
diagrams are more fundamental in the present case. The problem, of course,
is that the Dyson operator $\vec{T}$ is reordering the Heisenberg fields
only, while the $E_{\alpha \beta }$'s remain in their original order.

\subsection{Zero Dimensional QFT}

Let $a$ and $a^{\dagger }$ be annihilation operators for a single mode
harmonic oscillator, We have the commutation relations $\left[ a,a^{\dagger }%
\right] =1$ and $a\Omega =0$. Let us consider the observable $q=za^{\dagger
}+z^{\ast }a$ where $z$ is a complex number. The Baker-Campbell-Hausdorff
theorem says that $\exp \left\{ itq\right\} =\exp \left\{ itza^{\dagger
}\right\} \exp \left\{ -\frac{1}{2}|z|^{2}t^{2}\right\} \exp \left\{
itz^{\ast }a\right\} $ which here has the same content as the expansion $%
\left( \ref{sum-over-connected-Wick-diagrams}\right) $. We can use a
diagrammatic presentation based on two types of vertex: the creation type 
%
%
%
\setlength{\unitlength}{.1cm}
\begin{picture}(10,5)
\label{pical}

\put(0,0){\dashbox{0.5}(10,0){ }}

\thicklines
\put(5,0){\circle*{1}}
\put(0,0){\oval(10,10)[tr]}

\end{picture}
%
which has weight $z,$ and the annihilation type 
%
%
%
%
\setlength{\unitlength}{.1cm}
\begin{picture}(10,5)
\label{picb}

\put(0,0){\dashbox{0.5}(10,0){ }}

\thicklines

\put(5,0){\circle*{1}}
\put(10,0){\oval(10,10)[tl]}

\end{picture}
%
\ which has weight $z^{\ast }$. If we take a vacuum expectation of $\exp
\left\{ itq\right\} $ then we need only consider connected diagrams having
no external lines - and there is only the one! The cumulant expansion is then

\begin{center}
$\left\langle \Omega |\exp \left\{ itq\right\} \Omega \right\rangle =\exp \{%
\dfrac{\left( it\right) ^{2}}{2!}$%
%
%
%
%
\setlength{\unitlength}{.1cm}
\begin{picture}(20,5)
\label{picc}

\put(0,0){\dashbox{0.5}(20,0){ }}

\thicklines

\put(5,0){\circle*{1}}
\put(15,0){\circle*{1}}
\put(10,0){\oval(10,10)[t]}

\end{picture}
%
$\}\equiv \exp \left\{ -\frac{1}{2}t^{2}|z|^{2}\right\} $,
\end{center}

\noindent and we see that $q$ is Gaussian in the vacuum state. As is well
known, the odd moments vanish while the even moments are $\left\langle
\Omega |q^{2n}\Omega \right\rangle =|z|^{2n}\frac{\left( 2n\right) !}{2^{n}n!%
}$ where the combinatorial factor counts the number of ways to partition the 
$2n$ (time-ordered) vertices into $n$ (contraction) pairs. For instance, the
fourth moment involves three disconnected diagrams (our convention is that
we only consider the contractions as edges; the thin horizontal base line
does not affect connectivity!)

\bigskip

\begin{center}
$\left\langle q^{4}\right\rangle =$ 
%
%
%
\setlength{\unitlength}{.05cm}
\begin{picture}(35,5)
\label{picG1}

\put(0,0){\dashbox{0.5}(35,0){ }}

\thicklines

\put(5,0){\circle*{2}}
\put(15,0){\circle*{2}}
\put(20,0){\circle*{2}}
\put(30,0){\circle*{2}}
\put(10,0){\oval(10,10)[t]}
\put(25,0){\oval(10,10)[t]}
\end{picture}
%
+ 
%
%
%
\setlength{\unitlength}{.05cm}
\begin{picture}(30,5)
\label{picG2}

\put(0,0){\dashbox{0.5}(30,0){ }}

\thicklines

\put(5,0){\circle*{2}}
\put(10,0){\circle*{2}}
\put(20,0){\circle*{2}}
\put(25,0){\circle*{2}}
\put(15,0){\oval(10,10)[t]}
\put(15,0){\oval(20,20)[t]}
\end{picture}%
%
+ 
%
%
%
\setlength{\unitlength}{.05cm}
\begin{picture}(25,5)
\label{picG3}

\put(0,0){\dashbox{0.5}(25,0){ }}

\thicklines

\put(5,0){\circle*{2}}
\put(10,0){\circle*{2}}
\put(15,0){\circle*{2}}
\put(20,0){\circle*{2}}
\put(10,0){\oval(10,10)[t]}
\put(15,0){\oval(10,10)[t]}
\end{picture}
%
$=3|z|^{4}$.
\end{center}

\bigskip

We may also consider the variable $N=\left( a+z\right) ^{\dagger }\left(
a+z\right) =a^{\dagger }a+za^{\dagger }+z^{\ast }a+|z|^{2}$. We now
introduce two extra vertices: a scattering vertex 
%
%
%
%
\setlength{\unitlength}{.1cm}
\begin{picture}(10,5)
\label{picdo}

\put(0,0){\dashbox{0.5}(10,0){ }}

\thicklines

\put(5,0){\circle*{1}}
\put(0,0){\oval(10,10)[tr]}
\put(10,0){\oval(10,10)[tl]}

\end{picture}
%
\ with weight unity and a constant vertex 
%
%
%
%
%
\setlength{\unitlength}{.1cm}
\begin{picture}(10,5)
\label{picd}

\put(0,0){\dashbox{0.5}(10,0){ }}

\thicklines

\put(5,0){\circle*{1}}

\end{picture}
%
with weight $|z|^{2}$. To determine the vacuum expectation of $\exp \left\{
itN\right\} $, we once again sum over all connected diagrams with no
external lines. This gives

\bigskip

\begin{center}
$\left\langle \Omega |\exp \left\{ itN\right\} \Omega \right\rangle =\exp \{%
\dfrac{\left( it\right) }{1!}$%
%
%
%
%
%
\setlength{\unitlength}{.1cm}
\begin{picture}(10,5)
\label{piceb}

\put(0,0){\dashbox{0.2}(10,0){ }}

\thicklines

\put(5,0){\circle*{1}}

\end{picture}
%
+$\dfrac{\left( it\right) ^{2}}{2!}$%
%
%
%
%
\setlength{\unitlength}{.05cm}
\begin{picture}(20,5)
\label{picfb}

\put(0,0){\dashbox{0.5}(20,0){ }}

\thicklines

\put(5,0){\circle*{2}}
\put(15,0){\circle*{2}}
\put(10,0){\oval(10,10)[t]}

\end{picture}
%
+$\dfrac{\left( it\right) ^{3}}{3!}$%
%
%
%
%
\setlength{\unitlength}{.05cm}
\begin{picture}(30,5)
\label{picgb}

\put(0,0){\dashbox{0.5}(30,0){ }}

\thicklines

\put(5,0){\circle*{2}}
\put(15,0){\circle*{2}}
\put(10,0){\oval(10,10)[t]}
\put(25,0){\circle*{2}}
\put(20,0){\oval(10,10)[t]}

\end{picture}
%
$+\cdots \}$
\end{center}

\bigskip

The $n^{th}$ term in the exponential will look like 
%
%
%
\setlength{\unitlength}{.05cm}
\begin{picture}(70,5)
\label{pich}

\put(0,0){\dashbox{0.5}(37,0){ }}
\put(53,0){\dashbox{0.5}(17,0){ }}
\put(41,1){$\cdots$}
\thicklines

\put(5,0){\circle*{2}}
\put(15,0){\circle*{2}}
\put(25,0){\circle*{2}}
\put(35,0){\circle*{2}}
\put(55,0){\circle*{2}}
\put(65,0){\circle*{2}}
\put(10,0){\oval(10,10)[t]}
\put(20,0){\oval(10,10)[t]}
\put(30,0){\oval(10,10)[t]}
\put(40,0){\oval(10,10)[tl]}
\put(50,0){\oval(10,10)[tr]}
\put(60,0){\oval(10,10)[t]}

\end{picture}
%
and each such term has weight $|z|^{2}$ - since the scattering vertices all
have weight unity. All cumulants are equal and we therefore have 
\begin{equation*}
\left\langle \Omega |\exp \left\{ itN\right\} \Omega \right\rangle =\exp
\left\{ \sum_{n\geq 1}\dfrac{\left( it\right) ^{n}}{n!}|z|^{2}\right\} =\exp
\left\{ |z|^{2}\left( e^{it}-1\right) \right\}
\end{equation*}
and we recognize $N$ as having a Poisson distribution of intensity $|z|^{2}$%
. The moments of the variable $N$ are given as a polynomial of degree $n$ in 
$|z|^{2}$, vis. 
\begin{equation*}
\left\langle \Omega |N^{n}\Omega \right\rangle =\sum_{m=1}^{n}S\left(
n,m\right) |z|^{2m}
\end{equation*}
and, as is well-known in combinatorial analysis \cite{Riodain}, the
coefficients $S\left( n,m\right) $ are the Stirling numbers of the second
kind: they count the number of ways to partition $n$ items into $m$
non-empty subsets. To see why they arise here, consider the following
diagram contributing to $\left\langle N^{7}\right\rangle $:

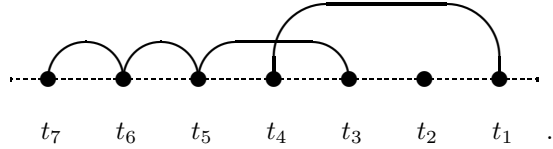
\begin{center}
%
%
%
%
%
\setlength{\unitlength}{.1cm}
\begin{picture}(70,20)
\label{picP}

\put(0,8){\dashbox{0.5}(70,0){ }}

\put(4,0){$t_7$}
\put(14,0){$t_6$}
\put(24,0){$t_5$}
\put(34,0){$t_4$}
\put(44,0){$t_3$}
\put(54,0){$t_2$}
\put(64,0){$t_1$}

\thicklines

\put(5,8){\circle*{2}}
\put(15,8){\circle*{2}}
\put(25,8){\circle*{2}}
\put(35,8){\circle*{2}}
\put(45,8){\circle*{2}}
\put(55,8){\circle*{2}}
\put(65,8){\circle*{2}}
\put(10,8){\oval(10,10)[t]}
\put(20,8){\oval(10,10)[t]}
\put(35,8){\oval(20,10)[t]}
\put(50,8){\oval(30,20)[t]}

\end{picture}
%
.
\end{center}

This diagram partitions the 7 vertices into 3 subsets, namely $\left\{
t_{7},t_{6},t_{5},t_{3}\right\} $, $\left\{ t_{4},t_{1}\right\} $ and $%
\left\{ t_{2}\right\} $, with each subset forming a connected sub-diagram.
This contributes $\left( |z|^{2}\right) ^{3}$ to $\left\langle
N^{7}\right\rangle $. Consulting a textbook on combinatorics to get the
Stirling numbers, we find $\left\langle N^{7}\right\rangle
=\sum_{m=1}^{7}S\left( 7,m\right) \left( |z|^{2}\right)
^{m}=|z|^{2}+63|z|^{4}+301|z|^{6}+350|z|^{8}+140|z|^{10}+21|z|^{12}+|z|^{14}$%
. Alternatively, we could draw all $B_{7}=$ 877 diagrams out! The numbers $%
B_{n}=\sum_{m=1}^{n}S\left( n,m\right) $ counting the total number of ways
to partition the $n$ vertices into non-empty subsets (of connected Goldstone
diagrams) are known as the Bell numbers.

\section{Quantum Stochastic Calculus}

Remarkably, the equivalence between expansions $\left( \ref
{sum-over-Goldstone-diagrams}\right) $\ and $\left( \ref
{sum-over-connected-Wick-diagrams}\right) $ is restored in the
non-commutative case in one very important situation. This is when we
consider the cases $\alpha $ and $\beta $ taking only the values $0,1$ in $%
\left( \ref{interaction}\right) $\ and when the two-point function $G$ is
replaced by a delta-function. Effectively the field is some form of quantum
white noise in time. As $\alpha ,\beta $ is restricted to either $0$ or $1$,
we shall have only four types of vertex: a constant vertex $E_{00}\otimes 1$%
, an emission vertex $E_{10}\otimes \Phi ^{\left( -\right) }$, an absorption
vertex $E_{01}\otimes \Phi ^{\left( +\right) }$ and a scattering vertex $%
E_{11}\otimes \Phi ^{\left( -\right) }\Phi ^{\left( +\right) }$.

The reason for the algebraic equivalence, despite the fact that the $%
E_{\alpha \beta }$ need not commute, is that many of the Goldstone diagrams
vanish identically. This is because to the singular nature of the two-point
function with respect to the simplicial integration in $\left( \ref
{Goldstone-term}\right) $. We note that absence of certain diagrams
describing moments of quantum noises has occurred elsewhere, in particular,
there is an elegant description of the various forms of independent quantum
processes in these terms \cite{Lenczewski}.

Let us introduce some specific notations \cite{Gough1}. We make the
replacements 
\begin{equation*}
\Phi _{t}^{\left( +\right) }\hookrightarrow \frak{a}_{t}\text{, }\Phi
_{t}^{\left( -\right) }\hookrightarrow \frak{a}_{t}^{\dagger }\text{, }%
G\left( t,s\right) \hookrightarrow \frak{g}\left( t-s\right) \text{,}
\end{equation*}
where $\frak{g}\left( t-s\right) =\kappa \frak{d}\left( t-s\right) +\kappa
^{\ast }\frak{d}\left( t-s\right) $. Here $\kappa $ is a complex damping
constant with $\gamma =2\func{Re}\left\{ \kappa \right\} >0$. The objects $%
\frak{d}_{\pm }\left( t-s\right) $ are one-sided delta functions defined by $%
\int f\left( s\right) \frak{d}_{\pm }\left( t-s\right) ds=f\left( t^{\pm
}\right) $%
\begin{equation}
\int_{-\infty }^{\infty }f\left( s\right) \frak{d}_{\pm }\left( s-t\right)
ds=\int_{-\infty }^{\infty }f\left( t+u\right) \frak{d}_{\pm }\left(
u\right) du=f\left( t^{\pm }\right) .
\end{equation}
Let us briefly indicate how to convert $U_{t}=\vec{T}\exp \left\{
-i\int_{0}^{t}\Upsilon _{s}ds\right\} $ to normal order \cite{Gough1} where $%
\Upsilon _{t}=E_{\alpha \beta }\otimes \left( \frak{a}_{t}^{\dagger }\right)
^{\alpha }\left( \frak{a}_{t}\right) ^{\beta }$ (we use a convention from
now on that repeated Greek indices are summed over values 0 and 1). When
evaluating Goldstone diagrams, we find that if the contractions are not
time-consecutive, that is, if we encounter $\frak{g}\left(
t_{i}-t_{j}\right) $ with $i>j+1$, then we force the multiple equalities $%
t_{i}=t_{i-1}=\cdots =t_{j+1}=t_{j}$ due to the time ordering, and so the
contribution vanishes. Only Goldstone diagrams with time-consecutive
contractions are non-zero.

Starting from the integro-differential equation $U_{t}=1-i\int_{0}^{t}%
\Upsilon _{s}U_{s}ds$, we have 
\begin{eqnarray*}
\left[ \frak{a}_{t}^{{}},U_{t}\right] &=&-i\int_{0}^{t}\left[ \frak{a}%
_{t}^{{}},\Upsilon _{s}\right] U_{s}ds \\
&=&-i\int_{0}^{t}\frak{g}\left( t-s\right) E_{1\beta }\left( \frak{a}%
_{t}^{{}}\right) ^{\beta }U_{s} \\
&=&-i\kappa E_{1\beta }\left( \frak{a}_{t}^{{}}\right) ^{\beta }U_{t}
\end{eqnarray*}
or $\frak{a}_{t}^{{}}U_{t}=\left( 1+i\kappa E_{11}\right) ^{-1}\left[ U_{t}%
\frak{a}_{t}^{{}}-i\kappa E_{10}U_{t}\right] $ and so 
\begin{equation}
\partial _{t}U_{t}=-iE_{\alpha \beta }\otimes \left( \frak{a}_{t}^{\dagger
}\right) ^{\alpha }\left( \frak{a}_{t}\right) ^{\beta }U_{t}\equiv \left( 
\frak{a}_{t}^{\dagger }\right) ^{\alpha }L_{\alpha \beta }U_{t}\left( \frak{a%
}_{t}\right) ^{\beta }  \label{white-noise-equation}
\end{equation}
where 
\begin{equation}
L_{\alpha \beta }=-iE_{\alpha \beta }-\kappa E_{\alpha 1}\frac{1}{1+i\kappa
E_{11}}E_{1\beta }.  \label{L's}
\end{equation}
We may interpret the conversion of the Schr\"{o}dinger equation to normal
ordered form as a change from a Stratonovich to an It\^{o} description. This
agrees with the interpretation given originally by von Waldenfels for
emission-absorption interactions \cite{vW1}.

Having normal-ordered the Schr\"{o}dinger equation, we now iterate to get 
\begin{eqnarray}
U_{t} &=&1+\int_{0}^{t}\left( \frak{a}_{s}^{\dagger }\right) ^{\alpha
}L_{\alpha \beta }U_{s}\left( \frak{a}_{s}\right) ^{\beta }ds  \notag \\
&=&\sum_{n\geq 0}\int_{\Delta _{n}\left( t\right) }\left( \frak{a}%
_{t_{n}}^{\dagger }\right) ^{\alpha _{n}}\cdots \left( \frak{a}%
_{t_{1}}^{\dagger }\right) ^{\alpha _{1}}\left( L_{\alpha _{n}\beta
_{n}}\cdots L_{\alpha _{1}\beta _{1}}\right) \left( \frak{a}_{t_{1}}\right)
^{\beta _{1}}\cdots \left( \frak{a}_{t_{n}}\right) ^{\beta _{n}}.  \notag \\
&&  \label{Ito-expansion}
\end{eqnarray}
This is reasonably familiar to quantum field theorists and such expressions
can be found for instance in Berezin's book \cite{Berezin}. If $f$ is a
suitable test function, we may consider its coherent (i.e. exponential)
vector $\left| \varepsilon \left( f\right) \right\rangle $ and take $\frak{a}%
_{t}\left| \varepsilon \left( f\right) \right\rangle =f\left( t\right)
\left| \varepsilon \left( f\right) \right\rangle $ and $\left\langle
\varepsilon \left( f\right) \right| \frak{a}_{t}^{\dagger }=\left\langle
\varepsilon \left( f\right) \right| f\left( t\right) ^{\ast }$. As $\left( 
\ref{Ito-expansion}\right) $ is normal ordered, we have no difficulty in
assigning a meaning to $\left\langle \varepsilon \left( f\right)
|U_{t}\varepsilon \left( g\right) \right\rangle $. At this stage we could
just as well take $\left( \ref{Ito-expansion}\right) $ as the definition of
the process, this is the starting point of the Maassen kernel calculus \cite
{LindsayMaassen}. As such the time-consecutive contraction property is built
into QSC, though in a way that is not readily apparent.

For the benefit of quantum probabilists, who may well be a little lost at
this stage, we convert $\left( \ref{white-noise-equation}\right) $ into more
familiar language \cite{HP}. Let $\Lambda _{t}^{\alpha \beta
}=\int_{0}^{t}\left( \frak{a}_{s}^{\dagger }\right) ^{\alpha }\left( \frak{a}%
_{s}\right) ^{\beta }ds$ and we interpret these as the four fundamental
quantum processes: $\Lambda _{t}^{00}$ is time, $\Lambda _{t}^{10}$ is
creation, $\Lambda _{t}^{01}$ is annihilation and $\Lambda _{t}^{11}$ is
conservation.\ Loosely speaking, we say $\left\{ X_{t}t\geq 0\right\} $ is
adapted if $\left[ \frak{a}_{s}^{\sharp },X_{t}\right] =0$ whenever $s>t$.
Setting $X_{t}^{\left( j\right) }=\int_{0}^{t}\left( \frak{a}_{s}^{\dagger
}\right) ^{\alpha }x_{\alpha \beta }^{\left( j\right) }\left( s\right)
\left( \frak{a}_{s}\right) ^{\beta }ds$ where the $x_{\alpha \beta }^{\left(
j\right) }\left( \cdot \right) $ are adapted, we see that putting to normal
order yields 
\begin{equation*}
X_{t}^{\left( 1\right) }X_{t}^{\left( 2\right) }=\int_{0}^{t}\left( \frak{a}%
_{s}^{\dagger }\right) ^{\alpha }\left[ X_{s}^{\left( 1\right) }x_{\alpha
\beta }^{\left( 2\right) }\left( s\right) +x_{\alpha \beta }^{\left(
1\right) }\left( s\right) X_{s}^{\left( 2\right) }+x_{\alpha 1}^{\left(
1\right) }\left( s\right) x_{1\beta }^{\left( 2\right) }\left( s\right) %
\right] \left( \frak{a}_{s}\right) ^{\beta }ds.
\end{equation*}
The basic idea goes back to Hudson and Streater \cite{HS}. In QSC, we
usually write $dX_{t}=x_{\alpha \beta }\left( t\right) d\Lambda _{t}^{\alpha
\beta }$ and the above result is presented as the quantum It\^{o} formula $%
d\left( X^{\left( 1\right) }X^{\left( 2\right) }\right) =X^{\left( 1\right)
}d\left( X^{\left( 2\right) }\right) +d\left( X^{\left( 1\right) }\right)
X^{\left( 2\right) }+d\left( X^{\left( 1\right) }\right) d\left( X^{\left(
2\right) }\right) $ along with the quantum It\^{o} table $d\Lambda
_{t}^{\alpha 1}d\Lambda _{t}^{1\beta }=d\Lambda _{t}^{\alpha \beta }$. The
equation $\left( \ref{white-noise-equation}\right) $ is then interpreted as
the It\^{o} quantum stochastic differential equation $dU_{t}=L_{\alpha \beta
}U_{t}d\Lambda _{t}^{\alpha \beta }$ with $U_{0}=1$. The coefficients
satisfy the identities $L_{\alpha \beta }+L_{\beta \alpha }^{\dagger
}+\gamma L_{1\alpha }^{\dagger }L_{1\beta }=0$ which are necessary and
sufficient for $U_{t}$ to be an adapted, unitary quantum stochastic process.
The formula for the product of several quantum integrals comes down to a
normal ordering problem which can ultimately be presented as a sum over
diagrams, or equivalently, a sum over partitions of the time indices: for
the classical case, see \cite{RotaWallstrom}.

\section{Markov Limits}

Finally, we wish to comment on how regular quantum fields can approximate
the singular fields considered above. Let $\lambda \neq 0$ be a parameter
and consider fields $\Phi _{t}^{\left( \pm \right) }\left( \lambda \right) $
with a regular two-point function $G_{\lambda }\left( \cdot \right) $ which
becomes a delta-function in the limit $\lambda \rightarrow 0$. In
particular, we may take $G\left( \cdot \right) $ to be a integrable function
with $\gamma =\int_{-\infty }^{\infty }G$ and $\kappa =\int_{0}^{\infty }G$
and assume that $G\left( -t\right) =G\left( t\right) ^{\ast }$. Then set $%
G_{\lambda }\left( t,s\right) =\lambda ^{-2}G\left( \dfrac{t-s}{\lambda ^{2}}%
\right) $. We would then argue that in the limit $G_{\lambda }$ converges to
the singular function $\frak{g}$ consider above. We consider the regular
unitary evolution operators 
\begin{equation*}
U_{t}\left( \lambda \right) =\vec{T}\exp \left\{ -i\int_{0}^{t}E_{\alpha
\beta }\otimes \left( \Phi _{s}^{\left( -\right) }\left( \lambda \right)
\right) ^{\alpha }\left( \Phi _{s}^{\left( +\right) }\left( \lambda \right)
\right) \right\}
\end{equation*}
and we claim that for bounded $E_{\alpha \beta }$, with $\left\| \kappa
E_{11}\right\| <1$, $U_{t}\left( \lambda \right) $ converges to the singular
process $U_{t}$ considered in the last section.

The first remark that we make is that the $\lambda \rightarrow 0$ limit
leads to the vanishing of each non-time-consecutive Goldstone diagram
contributing to $U_{t}\left( \lambda \right) $. Moreover, when the surviving
terms are computed and re-summed, we formally get the correct It\^{o}
expansion $\left( \ref{Ito-expansion}\right) $. Note that $L_{\alpha \beta
}=-iE_{\alpha \beta }-i\sum_{r=1}^{\infty }E_{\alpha 1}\left( -i\kappa
E_{11}\right) ^{r-1}E_{1\beta }$ giving the contribution to a
time-consecutive block with $\alpha $ outgoing, $\beta $ incoming lines and
a sum over $r$ successive scattering in between. We see that the condition $%
\left\| \kappa E_{11}\right\| <1$\ is necessary to sum the geometric series.

The re-summation is rather tedious, though it helps that we know what answer
to expect! We also have the issue of convergence, however, we settle this
below. We remark that it is sufficient to consider only the vacuum
convergence as the more general situation can be inferred from this when we
look at convergence in arbitrary but appropriately scaled coherent vector
states.

\subsection{Pul\'{e} Inequalities (Gaussian)}

Let us start with the case where we have emission and absorption only in the
interaction. The vacuum Goldstone diagrams, as we have seen in section 2,
consist of $n_{2}$, say, pair contractions only. A typical diagram, one of $%
\frac{\left( 2n_{2}\right) !}{2^{n_{2}}n_{2}!}$ having $2n_{2}$ vertices, is
sketched below for $n_{2}=6$:

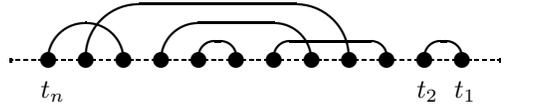
\begin{center}
%
%
%
%
\setlength{\unitlength}{.1cm}
\begin{picture}(70,20)
\label{picG4}

\put(0,5){\dashbox{0.5}(65,0){ }}

\thicklines

\put(5,5){\circle*{2}}
\put(10,5){\circle*{2}}
\put(15,5){\circle*{2}}
\put(20,5){\circle*{2}}
\put(25,5){\circle*{2}}
\put(30,5){\circle*{2}}
\put(35,5){\circle*{2}}
\put(40,5){\circle*{2}}
\put(45,5){\circle*{2}}
\put(50,5){\circle*{2}}
\put(55,5){\circle*{2}}
\put(60,5){\circle*{2}}

\put(10,5){\oval(10,10)[t]}
\put(57.5,5){\oval(5,5)[t]}
\put(42.5,5){\oval(15,5)[t]}
\put(27.5,5){\oval(5,5)[t]}
\put(30,5){\oval(20,10)[t]}
\put(27.5,5){\oval(35,15)[t]}
\put(59,0){$t_1$}
\put(54,0){$t_2$}
\put(4,0){$t_n$}
\end{picture}
%
.
\end{center}

There exists a permutation $\sigma $ of the $n=2n_{2}$ time indices which
re-orders to the\ diagram $D_{0}\left( n\right) $\ shown below

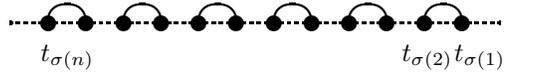
\begin{center}
%
%
%
%
%
\setlength{\unitlength}{.1cm}
\begin{picture}(70,20)
\label{picG5}

\put(0,5){\dashbox{0.5}(65,0){ }}

\thicklines

\put(5,5){\circle*{2}}
\put(10,5){\circle*{2}}
\put(15,5){\circle*{2}}
\put(20,5){\circle*{2}}
\put(25,5){\circle*{2}}
\put(30,5){\circle*{2}}
\put(35,5){\circle*{2}}
\put(40,5){\circle*{2}}
\put(45,5){\circle*{2}}
\put(50,5){\circle*{2}}
\put(55,5){\circle*{2}}
\put(60,5){\circle*{2}}

\put(47.5,5){\oval(5,5)[t]}
\put(57.5,5){\oval(5,5)[t]}
\put(37.5,5){\oval(5,5)[t]}
\put(27.5,5){\oval(5,5)[t]}
\put(17.5,5){\oval(5,5)[t]}
\put(7.5,5){\oval(5,5)[t]}

\put(59,0){$t_{\sigma (1)}$}
\put(52,0){$t_{\sigma (2)}$}
\put(4,0){$t_{\sigma (n)}$}
\end{picture}
%
.
\end{center}

The permutation is moreover unique if it has the induced ordering of the
emission times. Not all permutations arise this way, the ones that do are
termed admissible. We now consider an estimate of the $n-$th term in the
Dyson series: 
\begin{eqnarray*}
\sum_{D\in \mathcal{D}_{G}}\int_{\Delta _{n}\left( t\right) }\prod_{D}\left|
G_{\lambda }\right| &=&\sum_{\text{Admissible permutations}}\int_{\Delta
_{n}\left( t\right) }\prod_{D_{0}\left( n\right) }\left| G_{\lambda }\circ
\sigma \right| \\
&=&\int_{R\left( t\right) }\prod_{k=1}^{n_{2}}\left| G_{\lambda }\left(
t_{2k}-t_{2k-1}\right) \right|
\end{eqnarray*}
where $R\left( t\right) $ is the union of simplices $\left\{ \left(
t_{n},\cdots ,t_{1}\right) :t>t_{\sigma ^{-1}\left( n\right) }>\cdots
>t_{\sigma ^{-1}\left( 1\right) }>0\right\} $ over all admissible
permutations $\sigma $. $R\left( t\right) $ will be a subset of $\left[ 0,t%
\right] ^{2n_{2}}$ and if we introduce variables $t_{2k}$ and $%
s_{2k}=t_{2k}-t_{2k-1}$ for $k=1,\cdots ,n_{2}$ it is easily seen that the
above is majorized by $|\kappa |^{n_{2}}\times \dfrac{\max \left( t,1\right)
^{n_{2}}}{n_{2}!}$. This the Pul\`{e} inequality \cite{Pule} and\ clearly
gives the uniform absolute estimate required to sum the series.

\subsection{Pul\`{e} Inequalities (Poissonian)}

We now consider scattering, and constant, terms in the interaction \cite
{Gough7}. As we have seen there will be $B_{n}$ (the $n$-th Bell number)
Goldstone diagrams contributing to the $n-$th term in the vacuum Dyson
series expansion. The Bell numbers grow rapidly and have a complicated
asymptotic behaviour. The proliferation of diagrams is due to mainly to the
multiple scattering that now may take place.

Let us consider a typical Goldstone diagram. We shall assume that within the
diagram there are $n_{1}$ singleton vertices [$\cdots $%
%
%
%
%
%
\setlength{\unitlength}{.1cm}
\begin{picture}(10,5)
\label{picea}

\put(0,0){\dashbox{0.5}(10,0){ }}

\thicklines

\put(5,0){\circle*{1}}

\end{picture}
%
$\cdots ]$, $n_{2}$ contraction pairs [$\cdots $%
%
%
%
\setlength{\unitlength}{.05cm}
\begin{picture}(26,5)
\label{picfa}

\put(0,0){\dashbox{0.5}(7,0){ }}
\put(19,0){\dashbox{0.5}(7,0){ }}
\put(9,0){$\cdots$}
\thicklines

\put(5,0){\circle*{2}}
\put(21,0){\circle*{2}}
\put(13,0){\oval(16,16)[t]}

\end{picture}
%
$\cdots ]$, $n_{3}$ contraction triples [$\cdots $%
%
%
%
%
%
\setlength{\unitlength}{.05cm}
\begin{picture}(42,8)
\label{picga}

\put(0,0){\dashbox{0.5}(7,0){ }}
\put(19,0){\dashbox{0.5}(4,0){ }}
\put(35,0){\dashbox{0.5}(7,0){ }}
\put(9,0){$\cdots$}
\put(25,0){$\cdots$}
\thicklines

\put(5,0){\circle*{2}}
\put(21,0){\circle*{2}}
\put(37,0){\circle*{2}}

\put(13,0){\oval(16,16)[t]}
\put(29,0){\oval(16,16)[t]}

\end{picture}
%
$\cdots ]$, etc. That is the Goldstone diagram has a total of $%
n=\sum_{j}jn_{j}$ vertices which are partitioned into $m=\sum_{j}n_{j}$
connected subdiagrams. For instance, we might have an initial segment of a
diagram looking like the following:

\bigskip

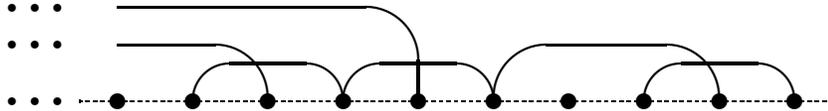
\begin{center}
%
%
%
%
%
%
%
%
%
%
\setlength{\unitlength}{.1cm}
\begin{picture}(110,20)
\label{picga}

\put(5,0){\dashbox{0.5}(100,0){ }}

\thicklines

\put(-4,0){\circle*{1}}
\put(-1,0){\circle*{1}}
\put(2,0){\circle*{1}}

\put(-4,7.5){\circle*{1}}
\put(-1,7.5){\circle*{1}}
\put(2,7.5){\circle*{1}}

\put(-4,12.5){\circle*{1}}
\put(-1,12.5){\circle*{1}}
\put(2,12.5){\circle*{1}}

\put(10,0){\circle*{2}}
\put(20,0){\circle*{2}}
\put(30,0){\circle*{2}}
\put(40,0){\circle*{2}}
\put(50,0){\circle*{2}}
\put(60,0){\circle*{2}}
\put(70,0){\circle*{2}}
\put(80,0){\circle*{2}}
\put(90,0){\circle*{2}}
\put(100,0){\circle*{2}}

\put(90,0){\oval(20,10)[t]}
\put(75,0){\oval(30,15)[t]}
\put(50,0){\oval(20,10)[t]}
\put(30,0){\oval(20,10)[t]}
\put(10,0){\oval(40,15)[tr]}
\put(10,0){\oval(80,25)[tr]}

\end{picture}
%
\end{center}

There will exist a permutation $\sigma $ of the $n$ vertices which will
reorder the vertices so that we have the singletons first, then the pair
contractions, then the triples, etc., so that we obtain a picture of the
following type

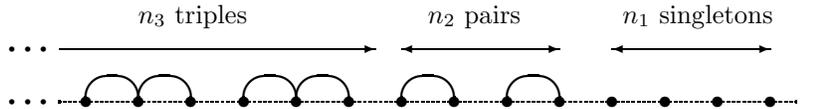
\begin{center}
%
%
%
%
%
%
%
%
%
%
%
%
%
%
\setlength{\unitlength}{.07cm}
\begin{picture}(150,20)
\label{picga}

\put(5,0){\dashbox{0.5}(140,0){ }}

\put(110,10){\vector(1,0){30}}
\put(140,10){\vector(-1,0){30}}
\put(112,15){$n_1$ singletons}

\put(70,10){\vector(1,0){30}}
\put(100,10){\vector(-1,0){30}}
\put(75,15){$n_2$ pairs}

\put(5,10){\vector(1,0){60}}
\put(20,15){$n_3$ triples}

\thicklines

\put(-4,0){\circle*{1}}
\put(-1,0){\circle*{1}}
\put(2,0){\circle*{1}}

\put(-4,10){\circle*{1}}
\put(-1,10){\circle*{1}}
\put(2,10){\circle*{1}}

\put(10,0){\circle*{2}}
\put(20,0){\circle*{2}}
\put(30,0){\circle*{2}}
\put(40,0){\circle*{2}}
\put(50,0){\circle*{2}}
\put(60,0){\circle*{2}}
\put(70,0){\circle*{2}}
\put(80,0){\circle*{2}}
\put(90,0){\circle*{2}}
\put(100,0){\circle*{2}}
\put(110,0){\circle*{2}}
\put(120,0){\circle*{2}}
\put(130,0){\circle*{2}}
\put(140,0){\circle*{2}}

\put(15,0){\oval(10,10)[t]}
\put(25,0){\oval(10,10)[t]}
\put(45,0){\oval(10,10)[t]}
\put(55,0){\oval(10,10)[t]}
\put(75,0){\oval(10,10)[t]}
\put(95,0){\oval(10,10)[t]}

\end{picture}
%
.
\end{center}

The permutation is again unique if we retain the induced ordering of the
first emission times for each connected block. We now wish to find an
uniform estimate for the $n$-th term in the Dyson series, we have 
\begin{equation}
\sum_{\text{Goldstone diagrams}}\int_{\Delta _{n}\left( t\right) }\prod
\left| G_{\lambda }\right| \times \text{``weights''}  \label{a}
\end{equation}
where the weights are the operator norms of various products of the type $%
E_{\alpha _{n}\beta _{n}}\cdots E_{\alpha _{1}\beta _{1}}$. In general, the
weight is bounded by 
\begin{equation*}
\left\| E_{11}\right\| ^{n_{1}+2n_{2}+3n_{3}+\cdots }\times
C^{n_{1}+n_{2}+n_{3}+\cdots }
\end{equation*}
where $C=\max_{\alpha \beta }\left\| E_{\alpha \beta }\right\| $. This is
because each connected diagram of $j$ vertices will typically have one
emission and one absorption, but $j-2$ scattering vertices. The Pul\`{e}
argument of rearranging the sum over diagrams into a single integral over a
region $R\left( t\right) $ of $\left[ 0,t\right] ^{n}$ again applies and by
similar reason we arrive at the upper bound for $\left( \ref{a}\right) $
this time of the type 
\begin{equation*}
\sum\nolimits_{n_{1},n_{2},n_{3},\cdots }^{\prime }\left\| \kappa
E_{11}\right\| ^{n_{1}+2n_{2}+3n_{3}+\cdots }\times
C^{n_{1}+n_{2}+n_{3}+\cdots }\times \frac{\max \left( t,1\right)
^{n_{1}+n_{2}+n_{3}+\cdots }}{n_{1}!n_{2}!n_{3}!\cdots }.
\end{equation*}
Here the sum is restricted so that $\sum_{j}jn_{j}=n$. An uniform estimate
for the entire series is then given by removing this restriction: 
\begin{equation*}
\Xi \left( A,B\right) =\sum_{n_{1},n_{2},n_{3},\cdots }\frac{\exp \left\{
\sum_{j}\left( Aj+B\right) n_{j}\right\} }{n_{1}!n_{2}!n_{3}!\cdots }
\end{equation*}
where $e^{A}=\left\| \kappa E_{11}\right\| $ and $e^{B}=C\max \left(
t,1\right) $. Again we use the trick to convert a sum of products into a
product of sums 
\begin{eqnarray*}
\Xi \left( A,B\right) &=&\sum_{n_{1},n_{2},n_{3},\cdots }\prod_{j}\frac{\exp
\left\{ \left( Aj+B\right) n_{j}\right\} }{n_{j!}}=\prod_{j}\sum_{n}\frac{%
\exp \left\{ \left( Aj+B\right) n\right\} }{n!} \\
&=&\prod_{j}\exp \left\{ e^{\left( Aj+B\right) }\right\} \\
&=&\exp \left\{ \sum_{j}e^{Aj}e^{B}\right\} \\
&=&\exp \left\{ \frac{e^{A+B}}{1-e^{A}}\right\} .
\end{eqnarray*}
where we need $e^{A}<1$ to sum the geometric series - this however, is
precisely our condition that $\left\| \kappa E_{11}\right\| <1$.

\section{Conclusions}

We have established a Markov limit in the sense of \cite{AFL} which we may
write as 
\begin{equation*}
\vec{T}\left\{ \exp -i\int_{0}^{t}E_{\alpha \beta }\otimes \left( \Phi
_{s}^{\left( -\right) }\right) ^{\alpha }\left( \Phi _{s}^{\left( +\right)
}\right) ^{\beta }ds\right\} \hookrightarrow \vec{T}\left\{ \exp
-i\int_{0}^{t}E_{\alpha \beta }\otimes \left( \frak{a}_{s}^{\dagger }\right)
^{\alpha }\left( \frak{a}_{s}\right) ^{\beta }ds\right\} .
\end{equation*}
On the left hand side we have an unitary which can be expanded as a normal
ordered expression of the quantum fields in terms of Goldstone diagrams. The
right hand side can be developed as an expansion over time-consecutive can
be understood as Hudson-Parthasarathy unitary quantum stochastic process. We
have shown the non-time-consecutive terms on the left hand side make a
negligible contribution in the Markovian limit. Interpreting Weyl order as
Stratonovich form and Wick order as It\^{o} form, the above result can be
considered as a non-commutative version of the Wong-Zakai limit theorem for
classical processes.

The same holds for Fermi fields, however, the proof is complicated because
we have to take the limit in matrix elements of appropriately scaled number
states \cite{GoughSobolev}. The same basic estimates suffice once more and
in the limit we end up with the same process except with the $\Lambda
^{\alpha \beta }$ now being Fermionic noises. As one might suspect, we have
to bother ourselves collecting factors of $-1$, and one would expect to
obtain the same result if we dealt with $q$-commutation relations\cite
{Bozejko}.

We remark that the time-ordered exponentials developed in \cite{Holevo}
differ from the notions presented here, as we are time-ordering quantum
white noises and not It\^{o} differentials, though they do arise in models
for Markov limits of discrete time systems \cite{AttalPautrat}.

Finally, we mention that we also have the convergence of the Heisenberg
dynamics $U_{t}\left( \lambda \right) ^{\dagger }\left( X\otimes 1\right)
U_{t}\left( \lambda \right) $ to $U_{t}^{\dagger }\left( X\otimes 1\right)
U_{t}$ \cite{Gough7}. This requires a slightly deeper analysis, however, the
basic estimates above are again at the heart of things. We invite the reader
to try and imagine the Goldstone diagram expansion of $U_{t}\left( \lambda
\right) ^{\dagger }\left( X\otimes 1\right) U_{t}\left( \lambda \right) $ to
get an idea of what is involved.

\section{Acknowledgments}

I wish to thank Professors Bo\d{z}ejko, Lenczewski, M\l otkowski,
Wysocza\~{n}ski and the local organisers of the 25 QP Conference in the
Mathematical Research and Conference Center, B\c{e}dlewo, Poland, for their
most kind hospitality during which these results were presented.

\end{document}